\documentclass[useAMS,usenatbib]{mn2e}
\usepackage{graphics,rotate}
\def\gtsim{\mathrel{\hbox{\rlap{\hbox{\lower4pt\hbox{$\sim$}}}\hbox{$>$}}}}
\def\lesssim{\mathrel{\hbox{\rlap{\hbox{\lower4pt\hbox{$\sim$}}}\hbox{$<$}}}}
\def\chandra{\protect{\it Chandra}}

\def\spose#1{\hbox to 0pt{#1\hss}}
\def\approxlt{\mathrel{\spose{\lower 3pt\hbox{$\sim$}}
        \raise 2.0pt\hbox{$<$}}}
\def\approxgt{\mathrel{\spose{\lower 3pt\hbox{$\sim$}}
        \raise 2.0pt\hbox{$>$}}}
\def\approxpropto{\mathrel{\spose{\lower 3pt\hbox{$\sim$}}
        \raise 2.0pt\hbox{$\propto$}}}
\mathchardef\twiddle="2218

\def\multleft#1{\hbox to size{\vbox {\halign {\lft{##}\cr #1}}\hfill}\par}
\def\multright#1{\hbox to size{\vbox {\halign {\rt{##}\cr #1}}\hfill}\par}

\def\degmark{^\circ}
\def\today{\ifcase\month\or January\or February\or March\or April\or May\or
      June\or July\or August\or September\or October\or November\or December\fi
      \space\number\day, \number\year}
\def\<{\thinspace}

\def\arcsec{{\rm\thinspace arcsec}}
\def\cm{{\rm\thinspace cm}}

\def\km{{\rm\thinspace km}}

\def\Mpc{{\rm\thinspace Mpc}}

\def\s{{\rm\thinspace s}}

\def\kmps{\hbox{$\km\s^{-1}\,$}}

\def\psqcm{\hbox{$\cm^{-2}\,$}}
\def\psqarcsec{\hbox{$\arcsec^{-2}\,$}}

\def\kmpspMpc{\hbox{$\kmps\Mpc^{-1}$}}

\title[Temperature and metallicity maps of the Perseus
cluster]{Chandra temperature and metallicity maps of the Perseus
cluster core}

\author[Schmidt, Fabian and Sanders]
       {R. W. Schmidt,\thanks{E-mail: rschmidt@ast.cam.ac.uk}
        A. C. Fabian and J. S. Sanders\\ 
        Institute of Astronomy, University of Cambridge, Madingley Road,
        Cambridge CB3 0HA}
\date{
      Received }

\pagerange{\pageref{firstpage}--\pageref{lastpage}}
\pubyear{}

\begin{document}

\maketitle

\label{firstpage}

\begin{abstract}

\noindent We present temperature and metallicity maps of the Perseus
cluster core obtained with the {\chandra} X-ray Observatory.  We find
an overall temperature rise from $\sim3.0$ keV in the core to
$\sim5.5$ keV at 120 kpc and a metallicity profile that rises slowly
from $\sim0.5$ solar to $\sim0.6$ solar inside 60 kpc, but drops to
$\sim0.4$ solar at 120 kpc. Spatially resolved spectroscopy in small
cells shows that the temperature distribution in the Perseus cluster
is not symmetrical. There is a wealth of structure in the temperature
map on scales of $\sim10$ arcsec (5.2 kpc) showing swirliness and a
temperature rise that coincides with a sudden surface brightness drop
in the X-ray image. We obtain a metallicity map of the Perseus cluster
core and find that the spectra extracted from the two central X-ray
holes as well as the western X-ray hole are best-fit by gas with
higher temperature and higher metallicity than is found in the
surroundings of the holes. A spectral deprojection analysis suggests,
however, that this is due to a projection effect; for the northern
X-ray hole we find tight limits on the presence of an isothermal
component in the X-ray hole, ruling out volume-filling X-ray gas with
temperatures below 11 keV at 3$\sigma$.

\end{abstract}


\section{Introduction}
\label{intro}

The Perseus cluster, Abell\,426, at a redshift of $z=0.0183$ or
distance about 100~Mpc is the the nearest high luminosity cluster with
a high central surface brightness \citep[e.g.,
][]{Fabian81,Allen01,Fabian94}. This makes it the brightest cluster in
the X-ray sky.

The first {\chandra} subarcsecond-resolution X-ray images of the
cluster core around the central dominant galaxy NGC\,1275 were
presented by \citet[][ F00]{Fabian00a}. Using X-ray colours F00 showed
that the temperature of the gas decreases from about 6.5~keV to 3~keV
inward to the central galaxy NGC\,1275, which is surrounded by a
spectacular low-ionization, emission line nebula \citep[][ see also
\citealt*{Conselice01}]{Lynds70}.

The nucleus powers the radio source 3C84 (Pedlar et al 1990) which has
structures on various scales. The 0.5 arcmin-sized radio lobes
coincide with holes in the soft X-ray emission
(\citealt{Boehringer93,McNamara96}; F00). {\chandra} has revealed that
the holes are not due to absorption and have X-ray bright rims, which
are cooler than the surrounding gas (F00). It was found that the rims
are not distinguishable as sharp features on a 3--7~keV image and are
therefore not shock features, contrary to the early prediction of
\citet*{Heinz98}, and are not expanding supersonically. The simplest
interpretation of the low surface brightness is that they are devoid
of X-ray gas and have pressure support from cosmic rays and magnetic
fields. However they may contain some hotter gas at the virial
temperature, or even above, of the cluster with the radio plasma
having a low filling factor. In this scenario the rims consist of
cooler gas which has been swept aside. The two outer holes, one of
which was previously known, were identified with recently-found spurs
of low-frequency radio emission by \citet{Blundell00}.

The core of the Perseus cluster offers the rare opportunity to study
the radio source/intracluster gas interaction on arcsecond scales. In
the present study we have carried out a spectral analysis of the
{\chandra} images which provides us detailed information about the
cluster physics. In Sect.~\ref{observations} we describe the
observations including the full 29.0 ks {\chandra} image of the
Perseus cluster core. In Sect.~\ref{specanalysis} we describe our
method of spectral analysis and show temperature and metallicity
profiles, as well as maps. In Sect.~\ref{specdeproj} we carry out a
spectral deprojection analysis of the cluster and search for the
presence of X-ray gas in the northern X-ray hole. In
Sect.~\ref{conclusions} we conclude with a summary and a discussion.

We use $H_0=50\kmpspMpc$ and $q_0=\frac{1}{2}$ throughout. Unless
otherwise stated, quoted error bars are 1\,$\sigma$ (68.3\%
confidence).

\section{Observations}
\label{observations}

{\chandra} observed the Perseus cluster on 1999 September 20, 1999
November 28 and 2000 January 31 for 5.3\,ks, 9\,ks and 24.7\,ks,
respectively. The early observation was carried out with the ACIS-I
detector. The energy resolution of this observation has been seriously
affected by the radiation damage suffered by the front-illuminated
ACIS chips, so that we only use it to produce images. The two later
observations were carried out with the ACIS-S detector. The target was
centred very close to the middle of the back-illuminated ACIS chip S3
(ACIS-S3). The detector temperature during these observations was -110
C. Inspection of the light curves of the observations showed that the
later, longer, observation had been affected by flares, so that we had
to clean the light curve. This reduced the exposure time usable for
spectroscopic analysis to 14.9\,ks.

\begin{figure*}
\resizebox{\textwidth}{!}{\includegraphics{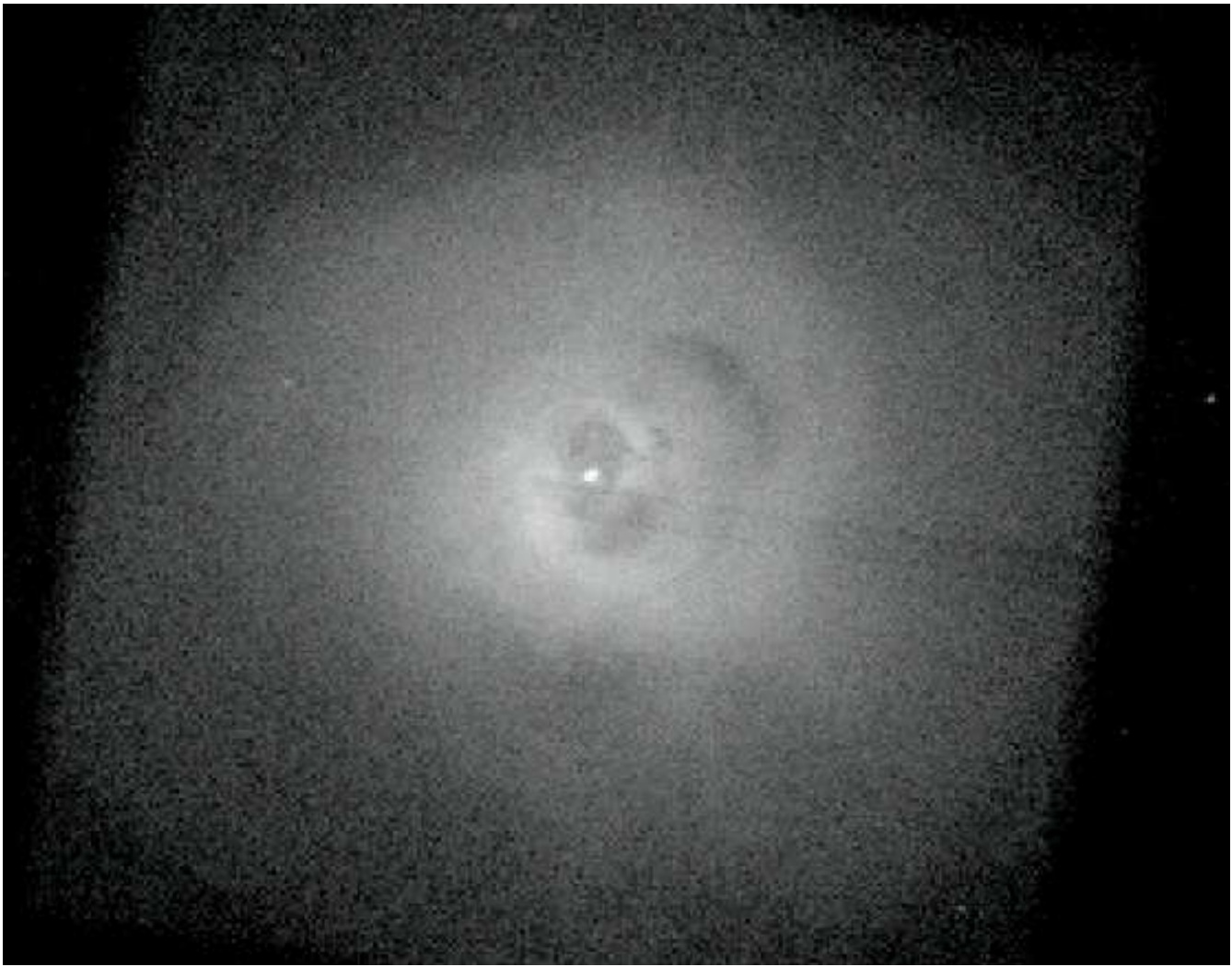}}
\caption{Combined 29.0 ks {\protect\chandra} image of the Perseus
  cluster core in the 0.5 to 7 keV band. The image size is 10.1 arcmin
  $\times$ 7.9 arcmin (312.5 kpc $\times$ 244.4 kpc). The linear
  boundaries visible at the edges of the frame are due to the three
  different orientations of the detector in the three added
  exposures.}
\label{combined}
\end{figure*}

In Fig.~\ref{combined} the combined 29.0\,ks image of all three
{\chandra} exposures of the Perseus cluster core is shown. The image
was constructed by concatenating the three event files using the {\sc
combine\_obsid} script available on the Chandra X-ray Observatory
Centre (CXC) web pages (www.cxc.edu).

The image was binned into (1.97 arcsec)$^2$ pixels. The well-known
features of the inner X-ray holes and the outer X-ray hole
(north-west) can be seen. In this stacked image a streak of emission
can be seen running through the southern hole (see also
\citealt{Fabian02}, hereafter F02). This feature was not obvious in
the image studied by F00 due to the dead column feature that runs
straight across it in the January 2000 image.  The southern outer
X-ray hole first discussed by F00 and another similar cavity
immediately to the west of it are also visible.

Between the western outer hole and the southern outer hole a curved,
elongated feature is apparent with a smaller count rate than in the
surrounding area. It appears to be a structure that connects the two
outer holes. It is reminiscent of the radio structure apparently
connecting the two outer radio lobes of the central radio source
3C~338 in Abell 2199 \citep{Giovannini98}.

The appearance of swirliness (F00, see also \citealt{Churazov00}) in
the outer regions of the image is created by a distinct surface
brightness drop that broadly coincides with a rise in the X-ray
temperature (with increasing radius), as we will show in the next
section.

\section{Spectral analysis}
\label{specanalysis}

\subsection{Calibration}
\label{basicmethod}

We extract spectra from regions of the {\chandra} data set using the
CIAO software package distributed by the CXC. We only work with
photons detected in the ACIS-S3 chip, which is also known as ACIS chip
7. We also restrict ourselves to the 0.5-7.0\,keV band in order to be
certain of a reliable energy calibration. Response matrix files ({\sc
RMF}) and ancillary response files ({\sc ARF}) were produced using the
calcrmf and calcarf programs by Alexey Vikhlinin available from the
CXC web pages.

Using the task {\sc grppha} from the {\sc FTOOLS} software package
provided by the NASA High Energy Astrophysics Science Archive Research
Center, the extracted spectra are grouped so that we have at least 20
counts in a bin corresponding to a certain range of energy
channels. This allows us to use Poisson error bars for spectral
fitting.

We have have generated background spectra using Markevitch's method
and program that is also available from the CXC web site. We extract
the background data from the same regions on the chip as the spectra
extracted from the data sets. In the case of the Perseus cluster core
the background in the ACIS-S3 chip in the 0.5-7.0\,keV band ($<\,1$
cts\,s$^{-1}$) is much smaller than the signal from the cluster
($83.4$ cts\,s$^{-1}$).

When fitting extracted spectra, we fit the November and the January
data set in parallel with the same spectral model, allowing only for a
different normalization of the two spectra. The different
normalizations are necessary because the exposures are affected by the
ACIS-S3 missing column feature in two different orientations. The
feature is due to 5 contiguous partially dead columns in the centre of
the chip that cause missing counts in an elongated region several
arcmin long and approximately 17\,{\arcsec} wide (due to the
spacecraft dither). This procedure affected only regions covering the
dead columns, in other regions the normalizations were consistent with
each other.

\subsection{Selection of regions}
\label{celldef}

In order to map the spectral properties of the cluster we divide the
cluster into cells with a constant number of cluster emission counts.
Since the background is not important for Perseus
(Sect.~\ref{basicmethod}) this is very similar to having an equal
number of counts in each cell.

We determine the cell shapes by iteratively dividing annular segments
with inner and outer radii $r_1$, and r$_2$ and bounding azimuthal
angles $\alpha_1$,$\alpha_2$ into regions with equal number of counts
(corrected for the background contribution as determined from
Markevitch's background files). Beginning with a full circle with
radius $r$ ($r_1=0$, $r_2=r$, $\alpha_1=0$, $\alpha_2=2\,\pi$) the
iterations are either exclusively performed on the radius, yielding a
sequence of annuli, or interchangingly on radius and azimuth, yielding
a more complex web of cells.

We carry out the analysis on three scales: (1) Simple annuli with
$\sim\,$35000 counts each, (2) cells with $\sim\,$18000 counts and (3)
cells with $\sim500$ counts each. The cell divisions were determined
using the 14.9 ks exposure taken on 1999 November 28. The mentioned
count values in this and the following sections refer to this
exposure.

\subsection{Annular spectra}

We begin the spectral analysis by studying azimuthally averaged
spectra from annuli with a large ($\sim\,$35000) number of
counts. This will give an overview of the radial variation of the
ambient X-ray gas temperature in Perseus.

We use XSPEC \citep{Arnaud96} to fit the X-ray spectra. Errors are
determined with the XSPEC tasks {\sc error} and {\sc steppar}. We
model the X-ray emission from the intracluster gas using the {\sc
APEC} plasma emission code (version 1.10). It is available on the web
at hea-www.harvard.edu/APEC and is also included in the latest XSPEC
versions. We use this newer code instead of the traditional MEKAL
plasma code \citep{Kaastra93,Liedahl95} because it provides a
significant improvement of the spectral fits. For example, when
fitting a spectrum extracted from an annulus with inner radius
23\,{\arcsec} and outer radius 32\,{\arcsec} around the nucleus (542
degrees of freedom) we obtain a $\chi^2=625.7$ for the {\sc APEC}
model instead of $\chi^2=662.0$ for MEKAL. The change in best-fit
parameters, however, is small; the best-fit absorption and temperature
are identical for both plasma codes. The metallicity changes slighly,
in this particular case from 0.42$\pm0.03$ solar for MEKAL to
0.46$\pm0.03$ solar for APEC. A large part of the improvement is due
to a better representation of the blue wing of the Fe L complex.

\begin{figure}
\resizebox{\columnwidth}{!}{\includegraphics{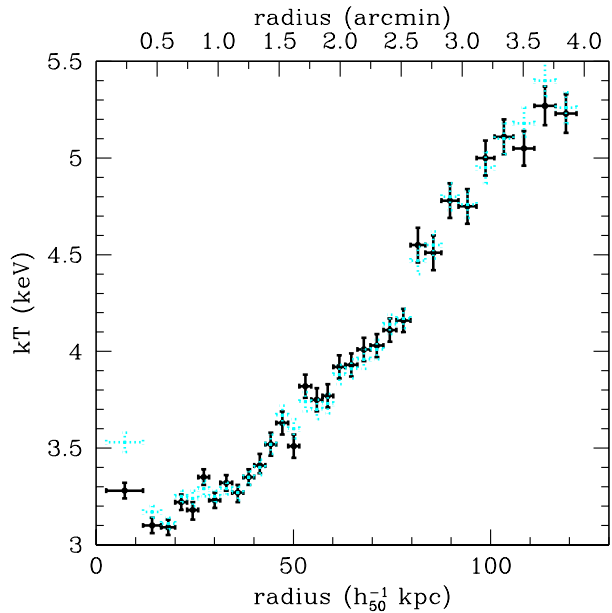}}
\caption{Temperatures from single temperature fits to annular spectra
between 2.5 and 121.8\,kpc (see also Figs.~\protect\ref{metal}
and~\protect\ref{absorb}). The temperatures with solid error bars have
been determined without fixing the galactic absorption (see
Fig.~\protect\ref{absorb}). For the dotted error bars the absorption
was fixed to the median of the fitted absorption values, N$_{\rm
H}$=0.143$\times10^{22}\psqcm$. Note the jump in the temperature
profile at a radius of 80\,kpc.}
\label{temp}
\end{figure}

\begin{figure}
\resizebox{\columnwidth}{!}{\includegraphics{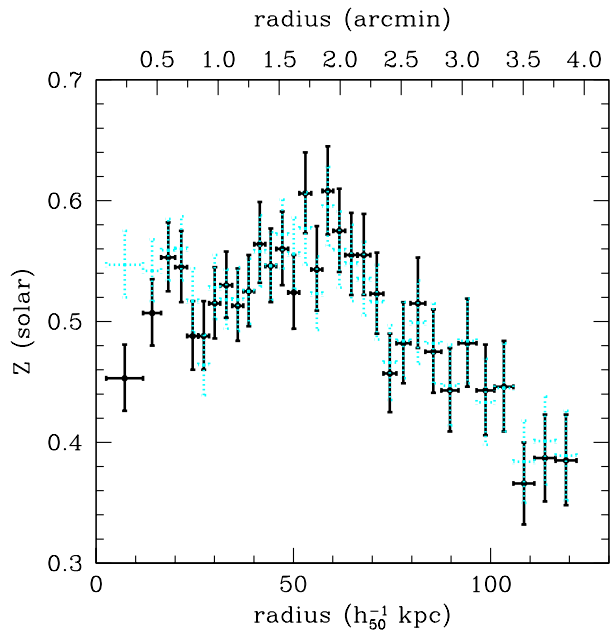}}
pwd
\caption{Metallicity profile from single temperature fits to annular
spectra between 2.5 and 121.8\,kpc (see also Figs.~\protect\ref{temp}
and~\protect\ref{absorb}). The metallicities with solid error bars
have been determined without fixing the galactic absorption
(Fig.~\protect\ref{absorb}). For the dotted error bars the absorption
was fixed to the median of the fitted absorption values, N$_{\rm
H}$=0.143$\times10^{22}\psqcm$.}
\label{metal}
\end{figure}

\begin{figure}
\resizebox{\columnwidth}{!}{\includegraphics{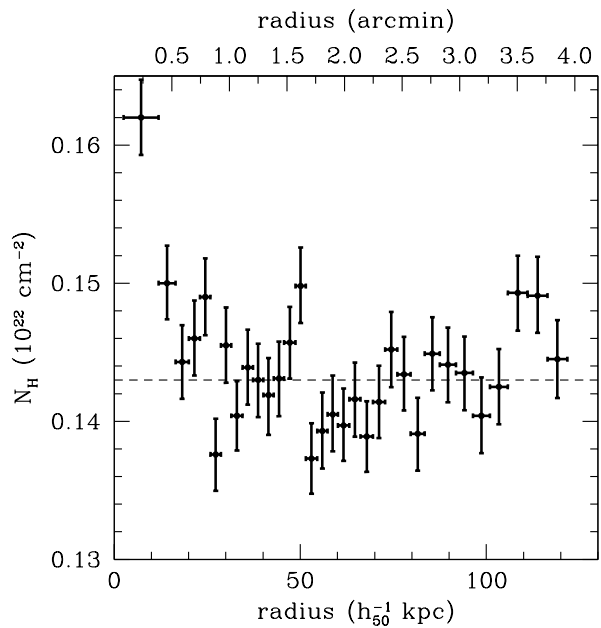}}
\caption{Column densities from single temperature fits to annular
spectra between 2.5 and 121.8\,kpc without fixing the absorbing
column density (see also Figs.~\protect\ref{temp}
and~\protect\ref{metal}). The dashed line indicates the median of
N$_{\rm H}$=0.143$\times10^{22}\psqcm$.}
\label{absorb}
\end{figure}

In order to model the emission from a single temperature we fit the
spectra with the following model:
\begin{equation}
{\rm MODEL}_1 = {\rm PHABS}\, (N_{\rm H}) \times {\rm APEC}\, (T,Z,K).
\label{apec}
\end{equation}
PHABS is the photoelectric absorption model by
\citet{Balucinska-Church92}. The free parameters equivalent hydrogen
column density $N_{\rm H}$, temperature $T$, average metallicity $Z$
(relative to the solar values according to \citealt{Anders89}) and
spectral normalization $K$ are given in brackets. In Figs.~\ref{temp},
\ref{metal} and~\ref{absorb} the results of the spectral fits for the
temperature, the metallicity and for the required photoelectric
absorption are shown. The fits have been done both with (dotted error
bars) and without (solid error bars) fixing the photoelectric
absorption. Except for the innermost annulus, the absorption is
consistent with being constant (Fig.~\ref{absorb}). For the fits with
fixed absorption we fixed it at the median of all fitted equivalent
hydrogen column densities $N_{\rm H}=0.143\times10^{22}\psqcm$ because
the Galactic value \citep{Dickey90} value from H$_{\rm\sc I}$ studies
($0.149\times10^{21}\psqcm$) appears to be slightly too high (see
Fig.~\ref{absorb}). In Tab.~\ref{anntable} we list the details of the
fits with free absorption. It can be seen from Figs.~\ref{temp} and
\ref{metal} that the best-fit values are very similar for models with
or without fixed absorption.

The temperature profile (Fig.~\ref{temp}) reveals a rapid temperature
decline from more than 5\,keV to about 3\,keV in the rims around the
inner X-ray holes. Due to the large number of counts, the error bars
are very small. There is a sudden jump in the temperature at a radius
of 80\,kpc, which coincides with the sudden brightness drop seen to
the north of the cluster centre (Fig.~\ref{combined}).

Note also that the temperature rises again in the core. In this
innermost ring, the nucleus and the inner 2.5\,kpc (4.8\,\arcsec)
around it have been cut out. It can be seen from Tab.~\ref{anntable}
that the spectral fit in this ring is not as good as the fits in most
of the other rings. This region encompasses most of the northern and
about half of the southern X-ray hole. It shows signs of excess
absorption (Fig.~\ref{absorb}) due to the infalling high-velocity
system \citep{Unger90,deYoung73,Briggs82} that was found in absorption
by {\chandra} (F00). A two-temperature model improves the fit and is
favoured by an F-test. The two-temperature fit yields a 4.6 keV gas
phase and a 1.9 keV gas phase, where the normalization of the
high-temperature phase is twice of the low-temperature phase. It only
improves the reduced ${\chi_{\nu}}^2$ to 1.34, however, so that this can
also not be viewed as an entirely acceptable fit. In general we find
that two-temperature fits do not provide a better description of the
annular spectra. Overall there is no convincing evidence for
multiphase gas in the {\chandra} spectra of the Perseus cluster.

In the metallicity profile (Fig.~\ref{metal}) a gradient from
$Z\sim0.6$ solar at a radius of 60\,kpc to $Z\sim0.4$ solar at a
radius of 120\,kpc is seen. There may be a peak of the metallicity
profile at $\sim50-60$ kpc, but the metallicities drop only slightly
inwards from 60\,kpc towards a central metallicity of $Z\sim0.5$
solar. Due to the large number of counts in the annuli, the relative
errors are normally less than 10 per cent so that the trend between
60\, and 120\,kpc is detected with high significance.

\begin{table*}
\caption{Single temperature fits to azimuthally averaged spectra. The
listed parameters are inner and outer radius ($r_1$, $r_2$),
temperature (column 3), metallicity (column 4) and equivalent
absorbing hydrogen column density (column 5). Describing the goodness
of fit we also list the $\chi^2$ value, the number of degrees of
freedom (DOF) and the reduced ${\chi_{\nu}}^2$ (columns 6 to 8).}
\label{anntable}
\begin{tabular}{@{}rrrrcccc@{}}
$r_1$ (kpc) & $r_2$ (kpc) & kT (keV) &
$Z$ (solar) & $N_{\rm H}$& $\chi^2$ & DOF & ${\chi_{\nu}}^2$\\
& & & & (10$^{21}$\,cm$^{-2}$)\\
\hline
2.5 & 11.9 & 3.28$\pm{0.04}$ & 0.45$\pm{0.03}$ & 1.62$\pm{0.03}$ & 746.7 & 541 & 1.38\\
11.9 & 16.5 & 3.10$\pm{0.04}$ & 0.51$\pm{0.03}$ & 1.50$\pm{0.03}$ & 593.0 & 542 & 1.09\\
16.5 & 20.0 & 3.09$\pm{0.04}$ & 0.55$\pm{0.03}$ & 1.44$\pm{0.03}$ & 597.6 & 531 & 1.13\\
20.0 & 23.1 & 3.22$\pm{0.04}$ & 0.55$\pm{0.03}$ & 1.46$\pm{0.03}$ & 607.3 & 533 & 1.14\\
23.1 & 25.9 & 3.18$^{+0.04}_{-0.05}$ & 0.49$\pm{0.03}$ & 1.49$\pm{0.03}$ & 541.7 & 532 & 1.02\\
25.9 & 28.7 & 3.35$\pm{0.04}$ & 0.49$\pm{0.03}$ & 1.38$\pm{0.03}$ & 679.1 & 549 & 1.24\\
28.7 & 31.5 & 3.23$\pm{0.04}$ & 0.52$\pm{0.03}$ & 1.45$\pm{0.03}$ & 636.9 & 540 & 1.18\\
31.5 & 34.5 & 3.32$\pm{0.04}$ & 0.53$\pm{0.03}$ & 1.40$\pm{0.03}$ & 714.8 & 565 & 1.27\\
34.5 & 37.3 & 3.27$^{+0.04}_{-0.05}$ & 0.51$\pm{0.03}$ & 1.44$\pm{0.03}$ & 724.3 & 539 & 1.34\\
37.3 & 40.1 & 3.35$\pm{0.04}$ & 0.53$\pm{0.03}$ & 1.43$\pm{0.03}$ & 620.3 & 545 & 1.14\\
40.1 & 42.9 & 3.41$^{+0.06}_{-0.04}$ & 0.56$^{+0.04}_{-0.03}$ & 1.42$\pm{0.03}$ & 649.6 & 550 & 1.18\\
42.9 & 45.7 & 3.52$\pm{0.06}$ & 0.55$\pm{0.03}$ & 1.43$\pm{0.03}$ & 587.8 & 544 & 1.08\\
45.7 & 48.7 & 3.63$\pm{0.06}$ & 0.56$\pm{0.03}$ & 1.46$\pm{0.03}$ & 635.0 & 565 & 1.12\\
48.7 & 51.5 & 3.51$\pm{0.06}$ & 0.52$\pm{0.03}$ & 1.50$\pm{0.03}$ & 588.0 & 548 & 1.07\\
51.5 & 54.5 & 3.82$\pm{0.06}$ & 0.61$\pm{0.03}$ & 1.37$\pm{0.03}$ & 678.2 & 573 & 1.18\\
54.5 & 57.3 & 3.75$\pm{0.06}$ & 0.54$^{+0.04}_{-0.03}$ & 1.39$\pm{0.03}$ & 622.1 & 552 & 1.13\\
57.3 & 60.1 & 3.77$\pm{0.06}$ & 0.61$\pm{0.04}$ & 1.41$\pm{0.03}$ & 592.3 & 550 & 1.08\\
60.1 & 63.2 & 3.92$\pm{0.06}$ & 0.57$^{+0.04}_{-0.03}$ & 1.40$\pm{0.03}$ & 617.9 & 574 & 1.08\\
63.2 & 66.2 & 3.93$\pm{0.06}$ & 0.56$^{+0.04}_{-0.03}$ & 1.42$\pm{0.03}$ & 740.8 & 570 & 1.30\\
66.2 & 69.5 & 4.01$\pm{0.06}$ & 0.56$\pm{0.03}$ & 1.39$\pm{0.03}$ & 763.0 & 579 & 1.32\\
69.5 & 72.8 & 4.03$\pm{0.06}$ & 0.52$\pm{0.03}$ & 1.41$\pm{0.03}$ & 634.5 & 571 & 1.11\\
72.8 & 76.1 & 4.11$\pm{0.06}$ & 0.46$\pm{0.03}$ & 1.45$\pm{0.03}$ & 534.2 & 569 & 0.94\\
76.1 & 79.7 & 4.16$\pm{0.06}$ & 0.48$\pm{0.03}$ & 1.43$\pm{0.03}$ & 724.3 & 577 & 1.26\\
79.7 & 83.5 & 4.55$\pm{0.09}$ & 0.52$\pm{0.04}$ & 1.39$\pm{0.03}$ & 692.1 & 589 & 1.18\\
83.5 & 87.5 & 4.51$\pm{0.09}$ & 0.47$^{+0.04}_{-0.03}$ & 1.45$\pm{0.03}$ & 651.1 & 596 & 1.09\\
87.5 & 91.8 & 4.78$\pm{0.09}$ & 0.44$^{+0.04}_{-0.03}$ & 1.44$\pm{0.03}$ & 666.4 & 598 & 1.11\\
91.8 & 96.4 & 4.75$\pm{0.09}$ & 0.48$\pm{0.04}$ & 1.43$\pm{0.03}$ & 688.0 & 597 & 1.15\\
96.4 & 101.0 & 5.00$\pm{0.09}$ & 0.44$\pm{0.04}$ & 1.40$\pm{0.03}$ & 698.2 & 592 & 1.18\\
101.0 & 105.8 & 5.11$\pm{0.09}$ & 0.45$\pm{0.04}$ & 1.42$\pm{0.03}$ & 659.4 & 603 & 1.09\\
105.8 & 111.1 & 5.05$\pm{0.09}$ & 0.37$\pm{0.03}$ & 1.49$\pm{0.03}$ & 686.9 & 612 & 1.12\\
111.1 & 116.4 & 5.27$\pm{0.10}$ & 0.39$\pm{0.04}$ & 1.49$\pm{0.03}$ & 663.2 & 608 & 1.09\\
116.4 & 121.8 & 5.23$\pm{0.10}$ & 0.39$\pm{0.04}$ & 1.44$\pm{0.03}$ & 664.3 & 592 & 1.12\\
\end{tabular}
\end{table*}

\subsection{Spectra in large cells}
\label{largecells}

\begin{figure*}
\protect\resizebox{5.8cm}{!}
{\includegraphics{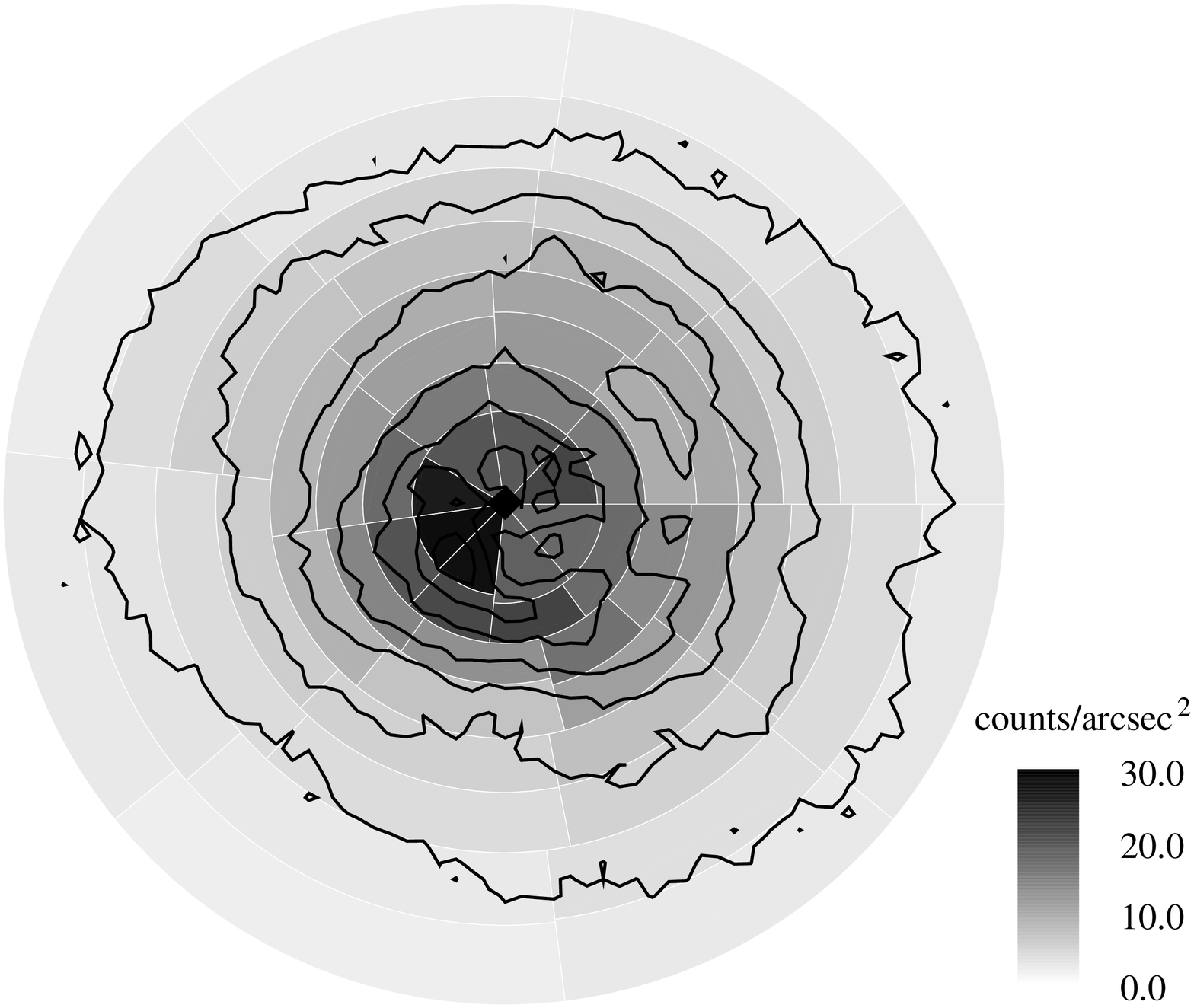}}
\protect\resizebox{5.8cm}{!}
{\includegraphics{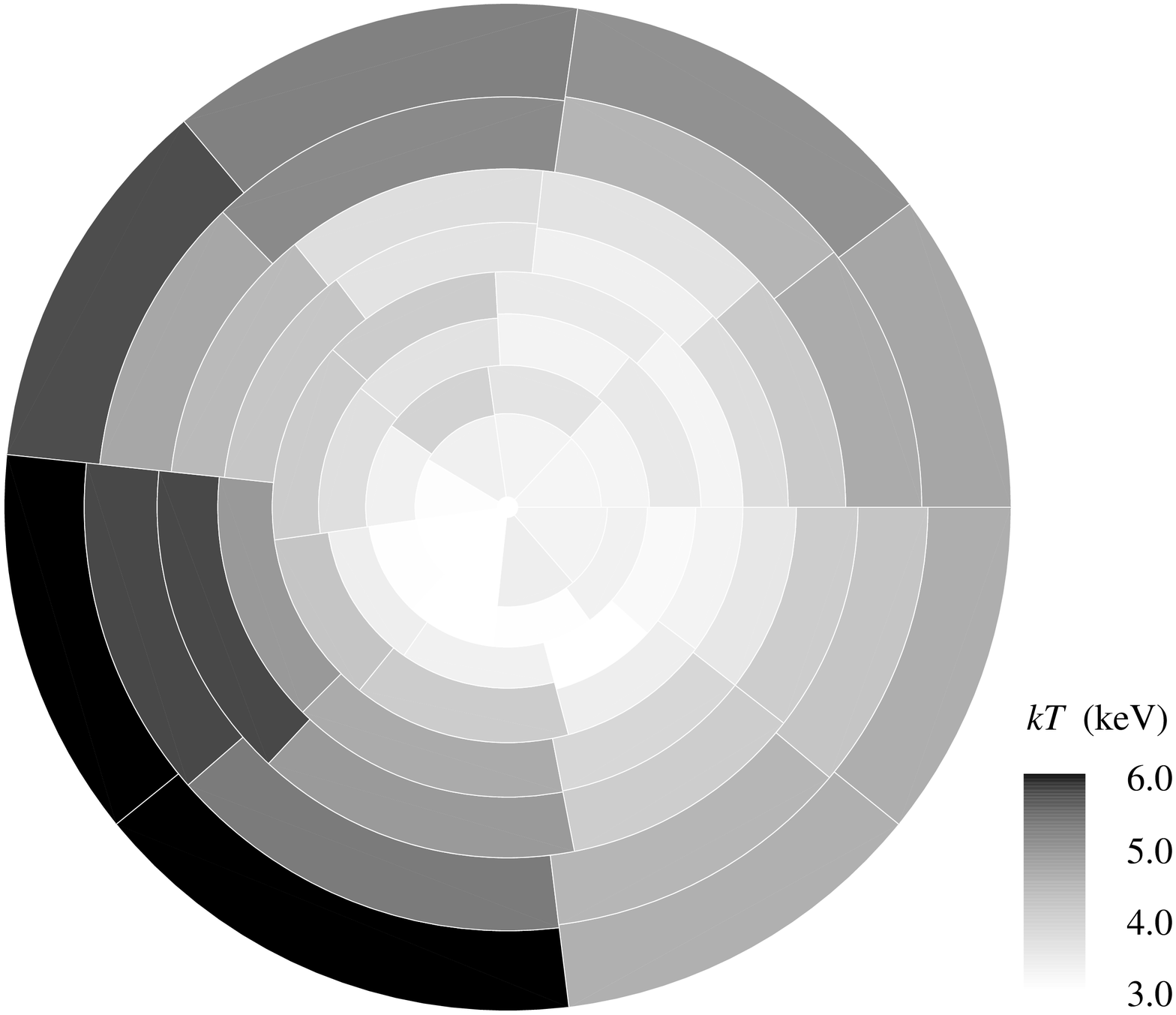}}
\protect\resizebox{5.8cm}{!}
{\includegraphics{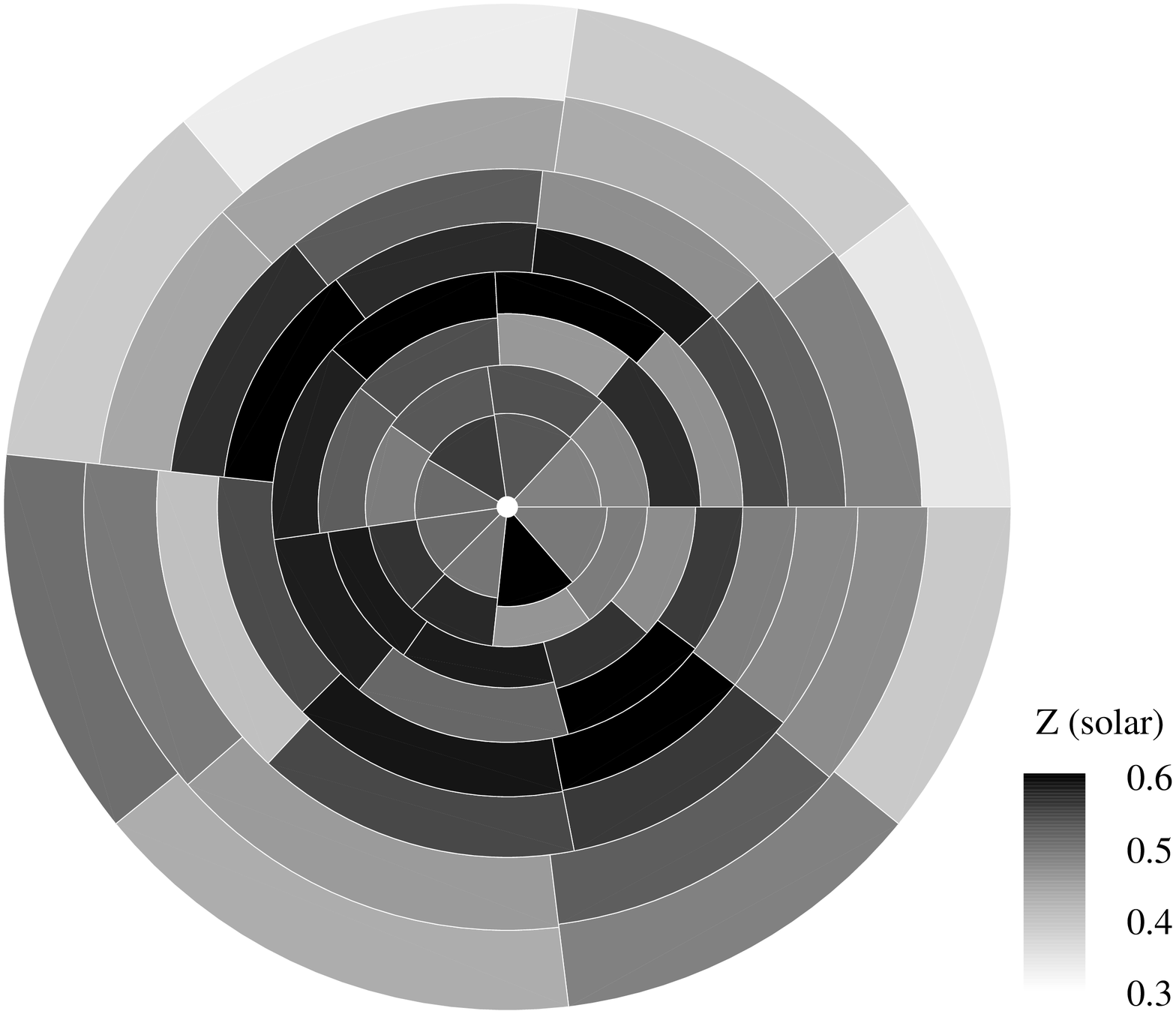}}
\caption{Intensity (left), temperature (middle) and metallicity
  (right) maps of the Perseus cluster core with $\sim18000$ counts per
  cell. The intensity map has been overlaid with contours at 3.7, 6.6,
  10.3, 14.8, 20.2, 26.4, 33.4 counts\,\psqarcsec. The counts were
  determined using the 14.9 ks exposure (see Sect.~\ref{celldef}). The
  diameter of the circle is 248.6\,kpc.}
\label{18k}
\end{figure*}

Following on to the azimuthally averaged spectra we next present the
results from cells with $\sim18000$ counts. In the left panel of
Fig.~\ref{18k} the definition of the cells and the intensity map of
the cluster with this resolution are shown. The contours of the
{\chandra} image are overlaid on top. The middle and right panels are
maps of the temperature and the average metallicity determined using
the single temperature model in eq.~(\ref{apec}) with free Galactic
absorption. The statistical uncertainty of the temperatures is less
than 0.1\,keV, the uncertainty of the metallicities is less than 10
per cent.

It was shown by F00 using X-ray colours that the temperature
distribution in the region is not circularly symmetric. The middle
panel in Fig.~\ref{18k} illustrates the overall temperature
distribution obtained with the {\sc APEC} plasma model fits in a
region of $\sim125$\,kpc radius around the nucleus. The large
asymmetric region in the centre at a temperature of 3 to 4\,keV is
surrounded by hotter gas with temperatures up to 6\,keV. The cool
region is extended towards the north and the west, whereas in the
south-east hotter gas is found at smaller radii. A comparison with the
intensity contours in the left panel shows that the outer intensity
contour has a similar shape to the isotherms at that radius; the
inward (towards smaller radii) bend of the outer intensity contour to
the south east of the cluster centre coincides with a
high-temperature region. The spectra extracted from the regions
corresponding to the inner X-ray holes and the outer hole (see the
lighter cell the left panel of Fig.~\ref{18k}) in the west are
best-fit by gas at a higher temperature than is found in their
immediate surroundings. The same may be true for the less obvious
southern X-ray hole discovered by F00 (see Fig.~\ref{combined}), but
here the distinction from the hotter gas that intrudes from the
south-east is not clear.

In the metallicity map (Fig.~\ref{18k}, right panel) small variations
of the metallicities are apparent between cells at similar radii. A
high-metallicity ring ($\sim0.6$ solar) around the cluster centre is
prominently seen at a radius of $\sim60\,$kpc where the peak in the
azimuthally averaged metallicity distribution in Fig.~\ref{metal} was
found. In addition to the higher temperature, the spectra extracted
from the inner X-ray holes and the western hole are also best-fit by
gas with a higher metallicity than is found in their immediate
surroundings.

We note that a comparison with the galaxy distribution in an optical
Digitial Sky Survey image of this region does not yield any obvious
correlation. Due to the relatively coarse scale of the metallicity
map, and due to projection effects, however, the question of
correlations between the metal distribution and the galaxy
distribution cannot be answered by this data set. Nevertheless, it is
exciting that the {\chandra} X-ray observations begin to approach
answering such questions.

\begin{figure}
\protect\resizebox{\columnwidth}{!}
{\includegraphics{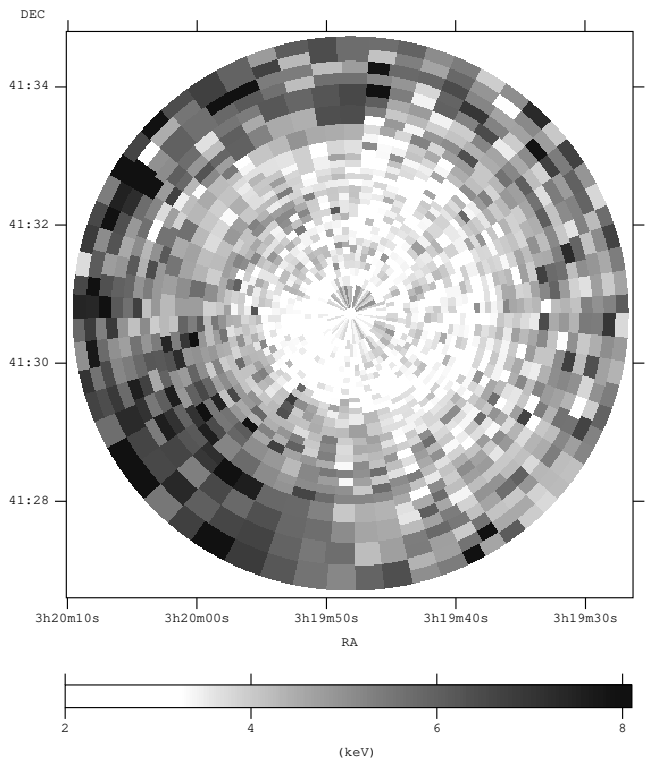}}
\caption{Temperature map of the perseus cluster core with 500
  counts/cell. The counts were determined using the 14.9 ks exposure
  (see Sect.~\ref{celldef}). The coordinates are J2000.}
\label{500counts}
\end{figure}

\subsection{Spectra in small cells}

In this section we finally use two grids of cells: firstly we
determine the amount of absorption and the metallicities in cells with
$\sim\,$9000 counts each using the model in eq.~\ref{apec}. We then
subdivide these cells further until we obtain a grid with $\sim500$
counts in each cell. In order to obtain cells that are not too
elongated, we limit the ratio
\begin{equation}
q=\frac{(\alpha_2-\alpha_1)\times(r_1+r_2)/2}{r_2-r_1}
\end{equation}
of cell width and depth (see Sect.~\ref{celldef}) to be greater than
$1/5$ and smaller than 5. The cells have a typical size (circle
section at half radius) of 10 arcsec (5.2 kpc). In these fine cells we
determine the temperature with a fixed metallicity at the value
determined in the larger cells. This procedure yields the finely
meshed temperature map shown in Fig.~\ref{500counts}. The relative
errors of the temperatures in this plot are of the order of $10-15$
per cent.

In order to obtain stable solutions for the absorption, we fix it to
the median value $N_{\rm H}=0.143\times10^{22}\psqcm$
(Fig.~\ref{absorb}) if it is smaller than $N_{\rm
  H}=0.16\times10^{22}\psqcm$, or otherwise fix it at the value
determined in the first step; in total, there were only two cells
where the local absorbing column densities of $N_{\rm
  H}=0.20\times10^{22}$ cm$^{-2}$ and $N_{\rm H}=0.21\times10^{22}$
cm$^{-2}$, respectively, had to be used. These two cells correspond to
the region where the infalling high-velocity system is seen in
absorption in the X-ray colour image by F00.

When looking at the goodness of the individual fits we noticed in a
small number of cases that the assumed (fixed) metallicity appeared to
underestimate the metallicity in these cells. In the cell with
$r_1=43.39$~kpc, $r_2=46.43$~kpc, $\alpha_1=346\degmark$ and
$\alpha_2=351\degmark$ (angles given in degrees, counterclockwise from
north), for example, we found a reduced ${\chi_{\nu}}^2$ value of 1.3
when the metallicity was fixed at $Z=0.434$ solar. The reduced
${\chi_{\nu}}^2$ drops to 1.1 for the best fit $Z=2.87^{+1.60}_{-1.74}$
solar. The constraint is entirely due to the iron L complex. This also
leads to an underestimate of the temperature:
$3.8^{+0.60}_{-0.56}$~keV compared $4.57^{+0.53}_{-0.77}$~keV if the
metallicity is left as a free parameter. Although the evidence is not
statistically compelling, this may be due to small-scale
inhomogeneities of the metallicity. However, currently the count rates
on small scales are too low and accordingly the uncertainties too
large to reliably determine metallicities on these scales.

In the temperature map the overall temperature distribution from
Fig.~\ref{18k} (middle panel) is again visible. In addition, the map
includes fine detail showing prominently that the temperature is
higher in the regions of the two inner and the north-western X-ray
holes than in their surroundings. Extending from the bright rim to the
south-east of the central X-ray holes the low-temperature (2 to 3 keV)
gas extends westwards and spirals around counter-clockwise. This
spiral shape of the colder gas corresponds to the swirly appearance of
the 0.5 to 7 keV emission shown in Fig.~\ref{combined}.  The shape of
lines of constant temperature at radii $\sim80$ kpc (2.6 arcmin)
discussed in Sect.~\ref{largecells} are also seen here to be similar
to the shape of the surface brightness drop in the X-ray emission. The
general spiral appearance of the cooler X-ray gas in the core was also
noted by F00 in their X-ray colour image; the shape of the temperature
distribution and the overall magnitude of the temperatures in the
cluster core as determined with X-ray colours by F00 are confirmed by
our individual spectral fits.

\section{Spectral deprojection}
\label{specdeproj}

\subsection{Azimuthally averaged profiles}

In the previous sections we have treated the cluster as a
two-dimensional object in order to learn about the overall spectral
properties. Since complicated projection effects must be present due
to the different locations of, for example, bubbles, brightness drops
or metals in the cluster, we attempt in this section to unravel some
of the projection effects that are present. However, this can only be
done by assuming a specific symmetry.

We deproject the observed spectra, in particular the temperature
profile (Fig.~\ref{temp}) and the metallicity profile
(Fig.~\ref{metal}), by assuming spherical symmetry and using the {\sc
\linebreak[4]projct} model available in XSPEC combined with our
standard {\sc APEC} model (eq.~\ref{apec}). In order to reduce the
number of parameters, only eight annuli with $\sim144,000$ counts each
were used (as measured in the January 2000 image).

The model assumes eight spherical, concentric shells filled with gas
with independent temperatures and metallicities. The projct model adds
the {\sc APEC} spectral contributions from the X-ray emission in the
shells in order to model each of the eight observed annular
spectra. All eight spectra are fitted at the same time\footnote{The
{\sc projct} model only deprojects a single data set at a time. We
thus used only the longest 14.9 ks data set from January 2000 for the
deprojection analysis reported in this section.}. Consistent with
Fig.~\ref{absorb}, we find no excess absorption except for the
innermost region. With the current model and due to the complicated
shape of the absorbing area (F00), however, it is not possible to
determine in which shell the absorbing material is located.

\begin{figure*}
\protect\resizebox{\columnwidth}{!}{\includegraphics{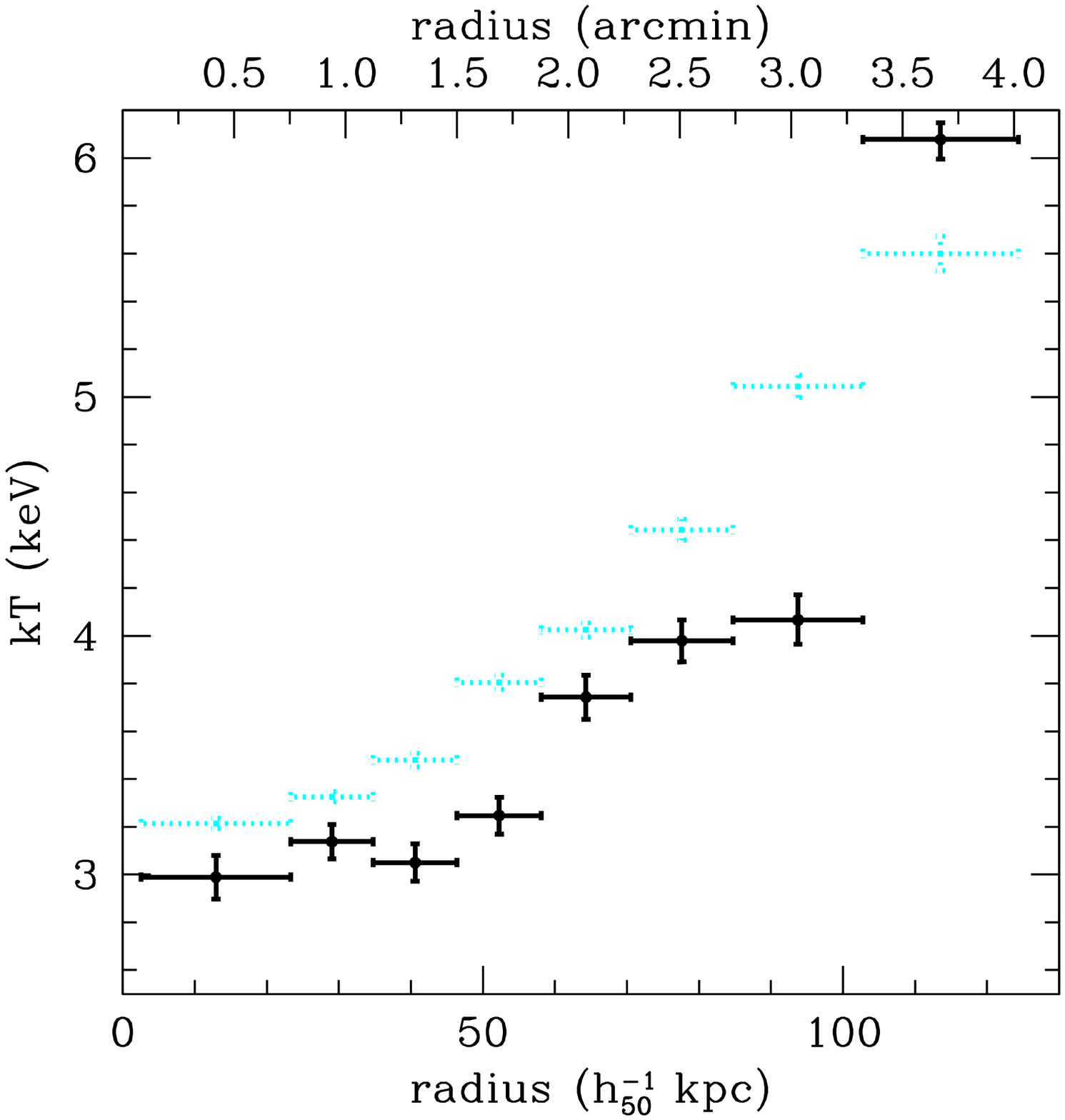}}
\hfill
\protect\resizebox{\columnwidth}{!}{\includegraphics{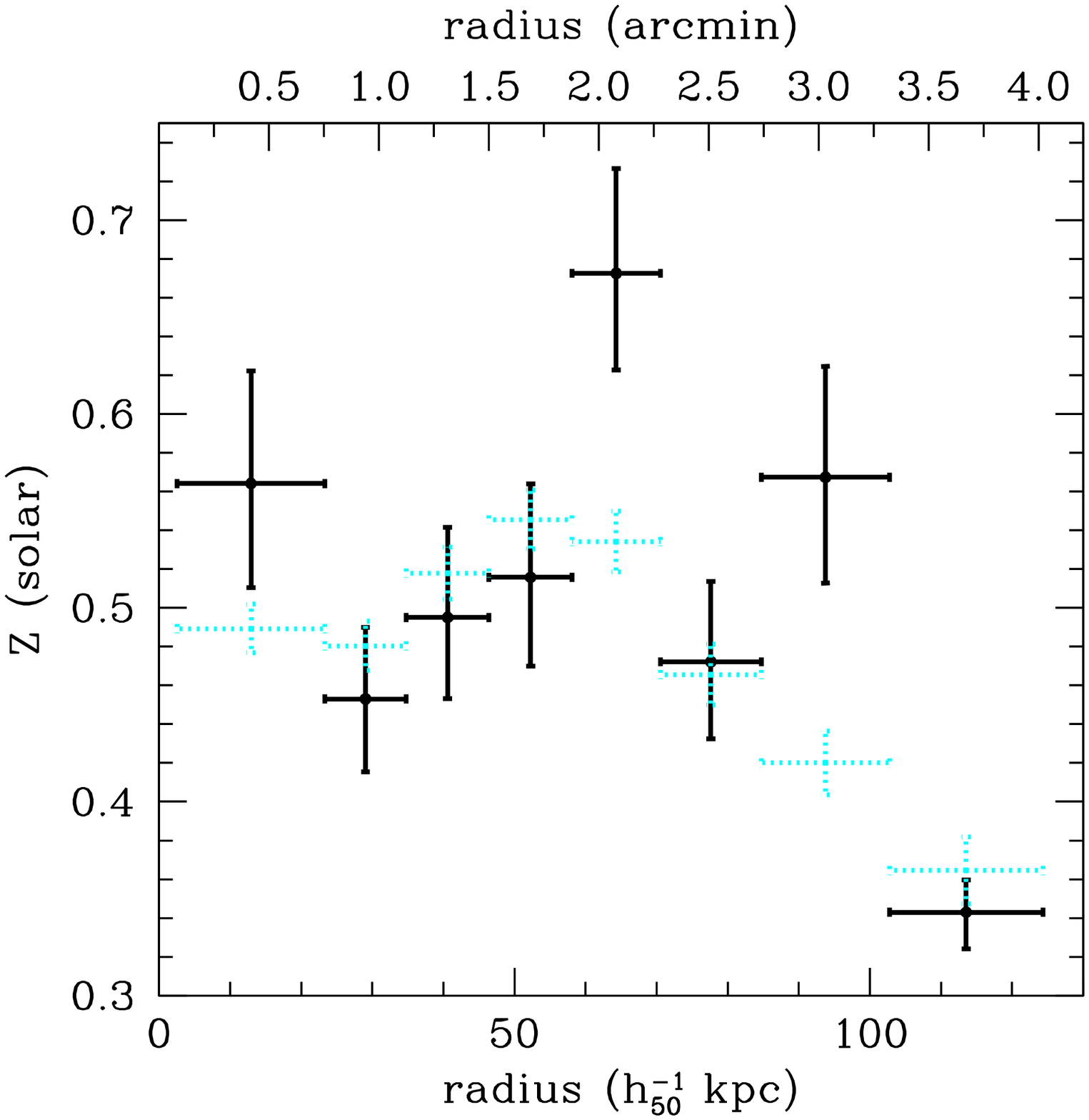}}
\caption{Deprojected temperature (left panel) and metallicity profiles
  (right panel) of the Perseus cluster core (solid lines). Also
  plotted with dotted lines are the corresponding {\it projected}
  quantities that are shown in Figs.~\ref{temp} and~\ref{metal} with
  higher resolution.}
\label{deproj}
\end{figure*}

The final deprojected temperature and metallicity profiles are shown
in Fig.~\ref{deproj}. We plot the deprojected values with solid error
bars, while the temperatures and metallicities determined from the
projected spectra are plotted with dotted error bars. Starting from
the outside, the temperature profile drops very quickly in the first
bin and continues to drop monotonically, but with a shallower slope
afterwards.  Also starting from the outside, the metallicity gradient
appears to be steeper down to $\sim70$ kpc than it was for the
projected spectra, but is very similar, albeit with larger error bars,
within 70 kpc. Inside a radius of $\sim70$ kpc, we find the
metallicity approximately constant at $Z\sim0.5$ solar.

\begin{figure}
\resizebox{\columnwidth}{!}{\includegraphics{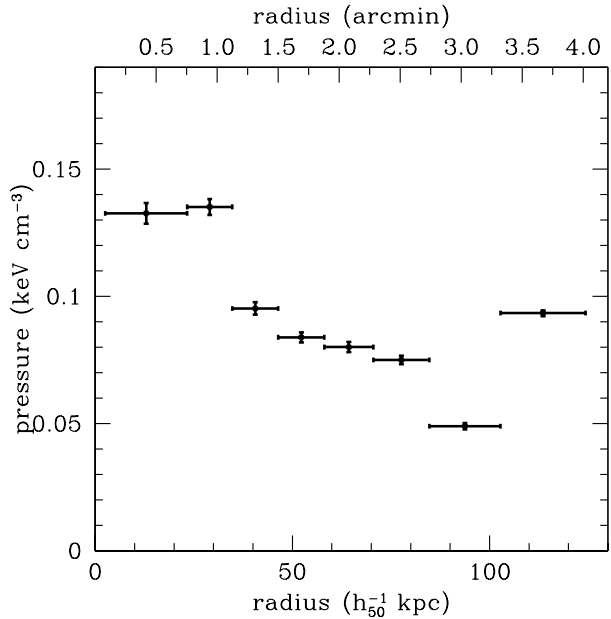}}
\caption{Electron pressure profile as determined using a spectral
  deprojection of the Perseus cluster core. The sudden rise in
  pressure in the outermost bin is not real, but an edge effect due to
  our data analysis (see text).}
\label{pressure}
\end{figure}

The X-ray data also allow us to determine the emission measure ${\rm
  EM}= n_{\rm e}^2\times V$ (via the normalization constant $K$ in
eq.~\ref{apec}), where $n_{\rm e}$ is the electron density and $V$ the
volume of the gas. Assuming that isothermal X-ray gas at the
deprojected temperature $T$ fills the spherical shells, we can
determine the pressure $p$ of the electrons, $p=n_{\rm e}\,$k$\,T$, as
shown in Fig.~\ref{pressure}. Note that the sudden rise in pressure in
the last bin is an edge effect due to our assumption that there is no
cluster emission outside this radius.  Similarly, the steep
deprojected temperature drop in the outermost bin of Fig.~\ref{deproj}
is an artifact since the cluster extends significantly beyond the
outermost radius accessible to the ACIS-S3 detector used. This is also
apparent from the fact that the deprojected temperature value is
best-fit higher than the projected value. The temperature, metallicity
and emission measure values further in, however, will not be affected
by this due to the steep brightness increase (see Fig.~\ref{18k}, left
panel) towards the nucleus.

\subsection{The X-ray emission from the X-ray holes}

In the projected temperature maps we found that the spectra extracted
from the two central X-ray holes as well as the western X-ray hole are
best-fit by hotter and more metal rich gas than their immediate
surroundings. It is possible that this is due to a projection effect,
whereby we would be seeing the hot gas above and below an empty
cavity.

In order to investigate the properties of the X-ray gas that fills the
region of the holes, we have carried a local deprojection analysis.
We have spectrally deprojected the emission emitted by 9 annular
segments (Sect.~\ref{celldef}) with azimuthal boundaries
$\alpha_1=350$ degrees and $\alpha_2=28$ degrees and 14000 counts per
segment (as measured in the January 2000 image) in the 0.5 to 7 keV
band. The innermost segment occupies the region of the northern inner
X-ray hole with radii between 5 and 27 arcsec of the nucleus.  The
analysis was analogous to the full-circle analysis in the last
section, except that the absorption was forced to be the same for all
segments and the metallicity in the inner hole was fixed at $Z=0.5$.
The best-fit absorption was $N_{\rm H}=1.40\times 10^{21}$ ${\rm
  cm}^{-2}$, the reduced $\chi^2$ of the best fit was
$\chi^2_{\nu}=1.14$.

\begin{figure}
\resizebox{\columnwidth}{!}{\includegraphics{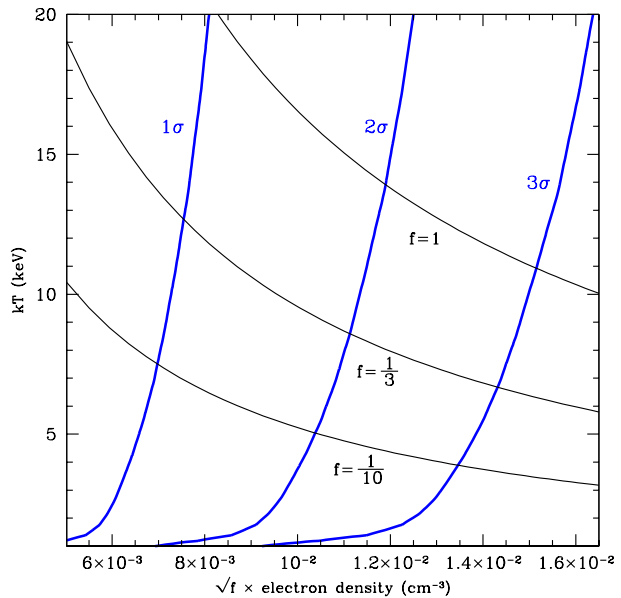}}
\caption{68\% ($1\;\sigma$), 95.4\%($2\;\sigma$) and
  99.73\%($3\;\sigma$) confidence contours (thick lines) on the
  presence of an isothermal APEC component in the northern inner X-ray
  hole. The hyperbolas (thin lines) denote the required parameter
  pairs for pressure equilibrium with the X-ray gas at larger radii,
  assuming filling factors $f=\frac{1}{10}$, $f=\frac{1}{3}$ and $f=1$
  of the X-ray gas.}
\label{holelimits}
\end{figure}

Interestingly, we find that almost the entire projected emission in
the innermost segment can be explained by the emission from the shells
above. In Fig.~\ref{holelimits} we show the allowed 68\%
($1\;\sigma$), 95.4\%($2\;\sigma$) and 99.73\%($3\;\sigma$) confidence
contours in the temperature/electron density plane of the parameters
for an APEC component in the innermost segment. It can be seen that
only a very low density plasma is allowed by the analysis. Since the
emission measure constrained by the observed X-ray data is
proportional to the volume of the gas, the allowed electron densities
depend on the filling factor $f$ of the X-ray gas in the hole. We thus
show the lines of pressure equilibrium with the second volume segment
for three different filling factors of the X-ray gas (the electron
pressure in the deprojected second segment is $p=n_{\rm e}\,T=0.17\,
{\rm keV\,cm}^{-3}$, slightly higher than the shell-averaged pressures
given in Fig.~\ref{pressure}). These data rule out at $3\sigma$ the
presence of X-ray gas at a temperature lower than 11 keV filling the
entire X-ray hole. At the same confidence level we can rule out
intracluster gas at the virial temperature of the cluster ($\sim$6.5
keV), filling one third of the X-ray hole.

\section{Discussion}
\label{conclusions}

We have presented in this paper the results from a detailed
temperature and metallicity mapping of the Perseus cluster core using
{\chandra} observations.

We work with the complete useful ACIS-S data set comprising a total of
23.9 ks. Beginning by averaging over large annuli around the nucleus,
we find that the temperature averaged in such annuli rises smoothly
from $\sim3.0$ keV to $\sim5.5$ keV at 120 kpc. A small kink is found,
however, in this profile at $\sim80$ kpc, which corresponds to a
surface brightness drop seen in the X-ray image. Using solar
abundances according to \citet{Anders89}, the metallicity rises from
$Z\sim0.4$ solar with decreasing radius to a maximum $Z\sim0.6$ solar
at $\sim60$ kpc. There may be a peak of the metallicity profile at
$\sim50-60$ kpc, which is reminiscent of such peaks in other clusters
\citep[e.g., ][]{Sanders01}. However, inside 60 kpc the metallicity
varies little, with a central metallicity of $Z\sim 0.5$ solar. We
have also carried out a spectral deprojection analysis of the radially
averaged profile, which yields a very similar picture.

Spatially resolved spectroscopy in small cells shows that the
temperature distribution in the Perseus cluster is not symmetrical. In
fact, the distribution of cold ($2-3$ keV) gas in the centre appears
to spiral outward and corresponds to the swirly appearance of the
X-ray emission (see Figs.~\ref{combined} and~\ref{500counts}). This
may suggest the presence of angular momentum of the intracluster gas
(F00). There are, however, other possibilities, such as the model by
\citet{Churazov00} (see also their adaptively smoothed ROSAT image)
who propose that rising radio bubbles (such as the X-ray holes) are
responsible for the overall spiral structure. The outer surface
brightness drop is traced by a rise in temperature (with increasing
radius). This is reminiscent, although not as large, of the
temperature drop at the cold front seen by \citet{Markevitch00}.  If
thermal conduction is an important effect in the intracluster medium,
such fronts could reflect the magnetic field structure.

A comparison of the metallicity distribution with the galaxy
distribution from an optical image from the Digital Sky Survey does
not show any obvious correlation between the galaxies and regions of
higher metallicity. Projection effects and the resolution of the
{\chandra} metallicity map, however, do not allow to rule out any
correlation yet.

We find that the spectra extracted from the two central X-ray holes as
well as the western X-ray hole are best-fit by hotter and more metal
rich gas than their immediate surroundings. Using a spectral
deprojection under the assumption of spherical symmetry in a sphere
segment containing the northern inner X-ray hole, we have tested
whether this could be due to a projection effect. Interestingly, we
find that most of the X-ray emission in the hole can be explained by
the projected emission of the shells further out. This directly
addresses the issue of the X-ray gas content of the X-ray holes
mentioned in the introduction; we find tight limits on the presence of
an isothermal component in the X-ray hole, ruling out volume-filling
X-ray gas with temperatures below 11 keV at 3$\sigma$.

The temperature distribution of the gas in the vicinity of the X-ray
holes is of great importance for theories of radio source heating to
counter the short cooling time of the X-ray gas in this cluster (e.g.,
F00). Longer exposure times are needed to reveal if there is cool gas
associated with the wake of the bubble, as, for example, in the model
by \citet{Churazov00}, who proposed that the radio bubbles would
transport cooler gas from the cluster centre `upwards'. The bubbles
may also transport or drag metals upwards. The subject of radio
bubbles and their interaction with the intracluster gas has recently
become the subject of much work
(\citealt*{Churazov01,Brueggen01,Quilis01,McNamara01}; F02). The
{\chandra} results provide essential observational input to test these
theories.

\section*{Acknowledgments}

We thank the anonymous referee for useful comments. ACF acknowledges
support by the Royal Society.



%
%

\bsp

\label{lastpage}

\end{document}